\documentclass[runningheads]{llncs}

\usepackage[utf8]{inputenc}
\usepackage{amssymb}
\usepackage{cite}
\usepackage{url}
\usepackage [strict=true, english = british]{csquotes}
\MakeOuterQuote{"}
\usepackage{listings}

\usepackage{amsmath}
\usepackage[dvipsnames]{xcolor}
\usepackage{tikz}
\usetikzlibrary{shapes,arrows,decorations.pathreplacing,positioning}

\usepackage{todonotes}
\usepackage{hyperref}
\usepackage[inline]{enumitem}

\newcommand{\MetagolPLS}{$\text{Metagol}_\text{PLS}$}

\begin{document}

\renewcommand\footnotemark{}
	
\title{Towards meta-interpretive learning of programming language semantics
}

\author{Sándor Bartha \and James Cheney}

\author{Sándor Bartha\inst{1} \and
	James Cheney\inst{1,2}}

%

%
	
\institute{	
	Laboratory for Foundations of Computer Science, University of Edinburgh\\
	\email{sandor.bartha@ed.ac.uk, jcheney@inf.ed.ac.uk}
\and
	The Alan Turing Institute
}	
	
\presetkeys{todonotes}{inline}{}

\maketitle

\begin{abstract}
	We introduce a new application for inductive logic programming: learning the semantics of programming languages from example evaluations. In this short paper, we explored a simplified task in this domain using the Metagol meta-interpretive learning system. We highlighted the challenging aspects of this scenario, including abstracting over function symbols, nonterminating examples, and learning non-observed predicates, and proposed extensions to Metagol helpful for overcoming these challenges, which may prove useful in other domains.
\end{abstract}

\section{Introduction}

Large systems often employ idiosyncratic domain specific languages, such as scripting, configuration, or query languages.
Often, these languages are specified in natural language, or no specification exists at all.
Lack of a clear specification leads to inconsistencies across implementations, maintenance problems, and security risks.
Moreover, a formal semantics is prerequisite to applying formal methods or static analysis to the language. 

In this short paper, we consider the problem: Given an opaque implementation of a programming language, can we reverse-engineer an interpretable semantics from input/output examples? 
The outlined objective is not merely of theoretical interest: it is a task currently done manually by experts.
\mbox{Krishnamurthi} et al.~\cite{Krishnamurthi:2019} cite a number of recent examples for languages such as  JavaScript, Python, and R that are the result of months of work by research groups. Reverse-engineering a formal specification involves writing a lot of small example programs, then testing their behaviour with an opaque implementation. 

Krishnamurthi et al.~\cite{Krishnamurthi:2019} highlights the importance of this research challenge. They describe the motivation behind learning the semantics of programming languages, and discuss three different techniques that they have attempted, showing that all of them have shortcomings.  However, inductive logic programming (ILP) was not one of the considered approaches.  A number of tools for computer-aided semantics exploration are already based on logic or relational programming, like $\lambda$Prolog~\cite{Miller:2012} or $\alpha$Prolog~\cite{Cheney:2008}, or PLT~Redex~\cite{Felleisen:2009}.  

Inductive logic programming seems like a natural fit for this domain:
it provides human-understandable programs, allows decomposing learning
problems by providing partial solutions as background knowledge (BK),
and naturally supports complex structures such as abstract syntax
trees and inference rules, which are the main ingredients of \emph{structural operational semantics} (SOS)~\cite{Plotkin:2004}.
These requirements make other popular learning paradigms, including
most statistical methods, hard to apply in this setting.

In this short paper we consider a simplified form of this task: given a base language, learn the rules for different extensions to the language from examples of input-output behavior.
We assume that representative examples of the language behaviour are available -- we are focusing on the learning part for now. 
We assume that we already have a parser for the language, and deal
with its abstract syntax
only. 
We also assume that the base language semantics (an untyped
lambda-calculus) is part of the background knowledge.

We investigated the applicability of meta-interpretive learning (MIL)~\cite{Muggleton:2014}, a state-of-the-art framework for ILP, on this problem. 
In particular we used Metagol~\cite{metagol}, an efficient implementation of MIL in Prolog. Our work is based on previous work on MIL~\cite{Andrew:2015}. We especially relied on the inspiring insight of how to learn higher-order logic functions with MIL~\cite{Cropper:2016}. Semantics learning is a challenging case study for Metagol, as interpreters are considerably more complex than the classic targets of ILP. 

We found that Metagol is not flexible enough to express the task of learning semantic rules from examples.
The main contribution of the paper is showing how to solve a textbook example of programming language learning by extending Metagol.  The extension, called \MetagolPLS, 
can handle learning scenarios with partially-defined predicates, can learn the definition of a single-step evaluation subroutine given only examples of a full evaluation, and can learn rules for predicates without examples and learn multiple rules or predicates from single examples.

We believe that these modifications could prove to be useful outside of the domain of learning semantics.
These modifications have already been incorporated to the main Metagol repository~\cite{metagol}.  We also discuss additional modifications, to handle learning rules with unknown function symbols and to handle non-terminating examples, which are included in \MetagolPLS but not Metagol.


 
All source code of \MetagolPLS\ and our semantics learning scenarios are available on GitHub: \url{https://github.com/barthasanyi/metagol_PLS}.

\section{A case study}\label{sec:case_study}

Due to space limits, we cannot provide a complete introduction to Metagol and have to rely on other publications describing it~\cite{Muggleton:2014}.  Briefly, in Metagol, an ILP problem is specified using examples, background knowledge (BK), and \emph{meta-rules} that describe possible rule structures, with unknown predicates abstracted as \emph{metavariables}.  Given a target predicate and examples, Metagol attempts to solve the positive examples using a meta-interpreter which may instantiate the meta-rules.  When this happens, the metarule instances are retained and become part of the candidate solution.  Negative examples are used to reject too-general candidate solutions.

First we give a formal definition of the general problem. Let $\mathcal{L}$ be the set of abstract syntax trees represented as Prolog terms. Let $L \subset \mathcal{L}$ be the language whose semantics we wish to learn, and let $V \subset L$ be the set of values (possible outputs). Let the behaviour of the opaque interpreter be represented as a function: $ I : L \to V \cup \{\bot\} $, where $\bot$ represents divergent computations. The function can be assumed to be the identity function on values: $\forall v \in V, I(v) = v$.  We do not have the definition of $I$, but we can evaluate it on any $e \in L$ term.

We assume that a partial model of the interpreter is defined in Prolog: let $B$ be the background knowledge, a set of Prolog clauses, which contains a partial definition of the binary $\mathtt{eval}$ predicate. We wish to extend the $\mathtt{eval}$ predicate so that it matches the $I$ function. Let $\mathcal{H}$ be the hypothesis space, a set of clauses that contains additional evaluation rules that may extend $B$.

The inputs are $L$, $I$, $B$ and $\mathcal{H}$. The expected output is $H \subset \mathcal{H}$, such that
\begin{enumerate}
	\item $ \forall e \in L, v \in V \,:\, I(e) = v \Longrightarrow B \cup H \vDash \mathtt{eval}(e,v) $
	\item $ \forall e \in L, v \in V \,:\, I(e) \ne v \Longrightarrow B \cup H \nvDash \mathtt{eval}(e,v) $
	\item $ \forall e \in L \,:\, I(e) = \bot \Longrightarrow \forall v \in V : B \cup H \nvDash \mathtt{eval}(e,v) $
\end{enumerate}

Note that in this learning scenario we cannot guarantee the correctness of the output, as we assumed that $I$ is opaque and we can only test its behaviour on a finite number of examples. We can merely empirically test the synthesized rules on suitable terms against the implementation, possibly adding terms to the examples where we get different results, and restarting the learning process. This actually matches the current practice by humans, as one reason for the tediousness of obtaining the semantics is that the existing implementation of the language is usually not intelligible.

As a case study of the applicability of Metagol to this general task, we chose a classic problem from PL semantics textbooks: extending the small-step structural operational semantics of the $\lambda$-calculus with pairs and its selector functions \lstinline|fst| and \lstinline|snd|. By analysing this problem we show how can we represent learning tasks in this domain with MIL, and what modifications of the framework are needed.

In this case the language $L$ contains $\lambda$-terms extended with pairs and selectors, and the background knowledge $B$ is an interpreter (SOS semantics) in Prolog implementing the $\lambda$-calculus:
\begin{lstlisting}
step(app(lam(X,T1),V),T2)   :- substitute(V,X,T1,T2).
step(app(T1,T2),app(T3,T2)) :- step(T1,T3).

eval(E1,E1) :- value(E1).
eval(E1,E3) :- step(E1,E2) , eval(E2,E3).
\end{lstlisting}
Here, \lstinline{substitute} is another BK predicate whose definition we omit, which performs capture-avoiding substitution.  The \lstinline{step} predicate defines a single evaluation step, e.g. substituting a value for a function parameter.  The \lstinline{value} predicate recognizes fully-evaluated values, and the \lstinline{eval} predicate either returns its first argument if it is a value, or evaluates it one step and then returns the result of further evaluation.

We wish to extend our calculus and its interpreter with pairs: a constructor \lstinline{pair} that creates a pair from two $\lambda$-terms, and two built-in operations: \lstinline|fst| and \lstinline|snd|, that extract the corresponding components from a pair. We want to learn all of the semantic rules that need to be added to our basic $\lambda$-calculus interpreter from example evaluations of terms that contain pairs. For example, we wish to learn that the components of the pair can be evaluated by a recursive call, and that a pair is a value if both of its components are values.

Our main contribution was interpreting this learning problem as a task
for ILP. We include the whole interpreter for the $\lambda$-calculus
in the BK. In MIL the semantic bias is expressed in the form of
meta-rules~\cite{Muggleton:2015}. Meta-rules are templates or schemes
for Prolog rules: they can contain predicate variables in place of
predicate symbols. We needed to write meta-rules that encompass the
possible forms of the small-step semantic rules required to evaluate
pairs.


	
Substitution is tricky on name binding operations, but fairly trivial on any other construct, and can be handled with a general recursive case for all such constructs. We assumed that we only learn language constructs that do not involve name binding, and included a full definition of substitution in the BK.

In general, we consider examples \lstinline{eval(e,v)} where \lstinline{e} is an expression and \lstinline{v} is the value it evaluates to (according to some opaque interpreter). Consider this positive example (Metagol's search is only guided by the positive examples):
\begin{lstlisting}
eval( app( lam(x,fst(var(x))) ,
           pair( app( lam(x,pair( app( lam( z, var(z)),var(x)),var(y))) , var(z)) , var(x)) ) ,
      pair(var(z),var(y)) )
\end{lstlisting}
which says that the lambda-term $(\lambda x.\mathtt{fst}(x))~((\lambda x.~(~(\lambda z. z) x, y))~z,x)$ evaluates to $(z,y)$. Using just this example, we might expect to learn rules such as:
\begin{lstlisting}
step(fst(pair(A,B)),A).
step(pair(A,B),pair(C,B)) :- step(A,C).
value(pair(A,B)) :- value(A),value(B).
\end{lstlisting}
The first rule extracts the first component of a pair; the second says that evaluation of a pair can proceed if the first subexpression can take an evaluation step.  The third rule says that a pair of values is a value.  Note that the example above does not mention \lstinline|snd|; additional examples are needed to learn its behavior.



Unfortunately, directly applying Metagol to this problem does not work.  
What are the limitations of the Metagol implementation that prevents it from solving our learning problem? We compared the task to the examples demonstrating the capabilities of Metagol in the official Metagol repository and the literature about MIL, and found three crucial features that are not covered:

\begin{enumerate}	
	\item For semantics learning, we do not know in advance what function symbols should be used in the meta-rules.  Metagol allows abstracting over predicate symbols in meta-rules, but not over function symbols.  
	\item 
Interpreters for Turing-complete languages may not halt.  Moreover, nontermination may give useful information about evaluation order, for example to distinguish lazy and eager evaluation.  Metagol does not handle learning nonterminating predicates.
	\item In semantics learning, we may only have examples for a relation \lstinline{eval} that describes the overall input/output behavior of the interpreter, but we wish to learn a subroutines such as \lstinline{value} that recognize when an expression is fully evaluated, and \lstinline{step} that describes how to perform one evaluation step.  Metagol considers a simple learning scenario with a single learned predicate with examples for that predicate.  
	
\end{enumerate}

In the following we investigate each difference, and show amendments to the Metagol framework that let us overcome them.


\section{Overview of \MetagolPLS}

\subsection{Function variables in the meta-rules}\label{sec:functions}

As a first-order language, Prolog does not allow variables in predicate or function positions of terms. 
The MIL framework uses predicate variables in meta-rules. In Metagol meta-rules can contain predicate variables because atomic formulas are automatically converted to a list format with the built-in \lstinline|=..| Prolog operator inside meta-rules. 

We demonstrated that function variables can be supported in a similar vein in the meta-interpretive learning framework, converting compound terms to lists inside the meta-rules. We added a simple syntactic transformation to \MetagolPLS to automate these conversions. 

As an example, consider a general rule that expresses the evaluation of the left component under a binary constructor. In this general rule for the fixed \lstinline{step} predicate there are no unknown predicates.
But we do not know the binary constructor of the abstract syntax of the language, which we wish to learn from examples.
With logic notation, we can write this general rule as the following:
\[\exists \mathbb{H} \quad \forall L1,L2,R : \mathtt{step}(\mathbb{H}(L1,R),\mathbb{H}(L2,R)) \leftarrow  \mathtt{step}(L1,L2)\]
where $\mathbb{H}$ stands for an arbitrary function symbol.
Using 
lists instead of compound terms, we can write this meta-rule in the following format:
\begin{lstlisting}
metarule(step2l,[step,H],([step,[H,L1,R],[H,L2,R]]
  :- [[step,L1,L2]])).
\end{lstlisting}





\subsection{Non-terminating examples}\label{sec:nonterm}

Interpreters for Turing-complete languages are inherently non-total: for some terms the evaluation may not terminate. Any learning method must be able to deal with non-termination, but due to the halting problem it is impossible to do exactly: any solution will be either unsound or incomplete.  Nevertheless, a pragmatic approach is to introduce some bound on the evaluation. We added a user definable, global depth limit to Metagol.  
By using this approach we lose some formal results about learnability, but it seems to work well in practice.

Non-termination 
can also distinguish lazy and eager evaluation strategies.
To able to separate the two evaluation strategies, we used a three-valued semantics for the examples. We distinguished non-termination from failure: in addition to the traditional  classification of the examples into positive and negative ones, we  introduced a third kind: non-terminating examples. 

A non-terminating example means that the evaluation exceeds the depth limit; positive or negative examples are intended to succeed or finitely fail within the depth limit.

\subsection{Non-observation predicate and multi-predicate learning}\label{sec:flexible}

Metagol learns one predicate, determined from the examples. The rules synthesized for this predicate can call predicates completely defined in the BK.  This is the usual \emph{single-predicate} and \emph{observation predicate learning} scenario.

In our task the examples are provided for the top level predicate: \lstinline{eval}, for which we do not want to learn new rules: it is defined in the BK. The semantic rules themselves that we want to learn are expressed by two predicates: \lstinline{step} and \lstinline{value}, called by the \lstinline{eval} predicate. The \lstinline{step} and \lstinline{value} predicates are partially defined in the BK: we have some predefined rules, but we want to learn new ones for the new language constructs.

We found that this more complex learning scenario can be expressed with interpreted predicates~\cite{Cropper:2016}. They have been used to learn higher order predicates; we show that they can also be used for \emph{non-observation predicate learning} and \emph{multi-predicate learning}.

We showed that interpreted predicates are useful for first order learning, too: as they are executed by the meta-interpreter, they may refer to predicates that are not completely defined in the BK, but need to be learnt. The meta-interpreter can simply switch back to learning mode from executing mode when it encounters a non-defined or partially defined predicate.

We added support for a special markup for predicate names to Metagol. We required the user to mark which predicates can be used in the head of a meta-rule, and similarly, to mark which predicates can be used in the body of a meta-rule.
This change extends the capabilities of Metagol in three ways:
\begin{enumerate}
	\item Non-observation predicate learning: We can include learned predicates in the BK, and learn predicates lower down in the call hierarchy. The examples can be for a predicate in the BK, and we can learn other predicates, that do not have their own examples.
	\item Multi-predicate learning: We can learn more than one predicate, and the examples can be for more than one predicate.
\end{enumerate}

This simple change nevertheless allows more flexible learning scenarios than the standard ILP setup. These changes have been incorporated into the official version of Metagol~\cite{metagol}.

\section{Evaluation}

Our modified version of Metagol and the tests are available on GitHub \url{https://github.com/barthasanyi/metagol_PLS}. All tests benefit from the changes that allow a more flexible learning scenario (Section~\ref{sec:flexible}),  learning non-terminating predicates (Section~\ref{sec:nonterm}), and function metavariables (Section~\ref{sec:functions}).



We coded three hand-crafted learning scenarios: learning the semantics of pairs, learning the semantics of lists (very similar to pairs), and learning the semantics of a conditional expression (\texttt{if then else}). Additionally we showed in a fourth scenario that we can distinguish eager and lazy evaluation of the $\lambda$-calculus based on a suitable term that terminates with lazy evaluation, but does not terminate with eager evaluation:

\begin{lstlisting}
eval(app(lam(x,var(y)), app(lam(x,app(var(x),var(x))),
                            lam(x,app(var(x),var(x))))),_)
\end{lstlisting}

All four case studies use the same hypothesis space (the same set of meta-rules), and the same BK. The meta-rules are similar to the one mentioned in Section \ref{sec:functions}. The BK contains the interpreter for the $\lambda$-calculus extended with simple integer arithmetic,  as well as two predicates that select a component. They are used in the induced rules for pairs, lists and conditionals:
\begin{lstlisting}
left(A,_,A).        right(_,B,B).
\end{lstlisting}

The evaluation examples are hand-crafted for each case study, and they are similar to the one showed earlier in Section \ref{sec:case_study}. The semantic rules are decomposed into multiple predicates in the output, since MIL tends to invent and re-use predicates.  We show this through the example of the synthesized semantics of conditionals. Conditionals are represented with two binary predicates in our target language: \mbox{\lstinline|if(A,thenelse(B,C))|}. We chose this format to avoid too many extra meta-rules for ternary predicates.

The induced rules for conditionals are (order re-arranged for readability):
\begin{lstlisting}
step(if(A,B),C) :- pred_1(A,B,C).        % Select apprpopriate branch
pred_1(false,A,B) :- pred_3(A,B).
pred_1(true,A,B) :- pred_2(A,B).
pred_2(thenelse(A,B),C) :- left(A,B,C).
pred_3(thenelse(A,B),C) :- right(A,B,C).
step(if(A,B),if(C,B)) :- step(A,C).      % Evaluate condition inside
value(false).                            % Boolean literals are values
value(true).                       
\end{lstlisting}


Finally, we demonstrated that the four learning tasks can be learned sequentially: we can learn a set of operational semantic rules from one task and add these to the BK for the next task. We chained all four demonstrations together, synthesizing a quite large set of semantic rules ($25$ rules total). Metagol does not scale up to learning this many rules in a single learning task: according to our preliminary investigations, the runtime is roughly exponential, which matches the theoretical results~\cite{Cropper:2018}. Even synthesizing half as many rules can take hours. Sequential learning have beenr implemented in Metagol~\cite{Lin:2014}, but the flexible learning scenarios required extending this functionality.

The examples run fairly fast: even the combined learning scenario finishes under $0.2$ seconds on our machine. However, during our preliminary experiments with hand-crafted examples we found that the running time of Metagol tasks greatly depends on the order of the examples: there can be orders of magnitude running time differences between example sets. Further research is needed to determine how to obtain good example sets.





\section{Conclusion and future work} 

This research is a first step towards a distant goal. Krishnamurthi et al.~\cite{Krishnamurthi:2019} make a strong case that the goal is both important and challenging.

We have demonstrated that with modifications MIL can synthesize structural semantic rules for a simple programming language from suitable (hand-crafted) examples. But we only considered relatively simple language semantics learning scenarios, so further work is need to scale up the method to realistic languages.  

The most crucial issue is scalability, which is the general problem for MIL. MIL does not scale well to many meta-rules and large programs. In our experiments we found that synthesizing less than $10$ rules is fast, but synthesizing more than $20$ seems to be impossible. As a comparison, the SOS semantics of real-world languages may contain hundreds of rules. Therefore we need a method to partition the task: to generate suitable examples that characterize the behaviour of the language on a small set of constructs, and to prune the set of meta-rules, which can be large. Our sequential learning case study ensures that once the problem is partitioned, we can learn the rules, but it does not help with the actual partitioning. Alternatively, other ILP systems that support learning recursive predicates, such as XHAIL~\cite{xhail} or ILASP~\cite{ilasp}, could be tried.

In our artificial example, substitution rules were added to the BK. In the presence of name binding constructs, correct (capture-avoiding) substitution is tricky to implement in Prolog. However, new language features sometimes involve name-binding and real languages sometimes employ non-standard definitions of substitution or binding.
Substitution, while ubiquitous, is a not a good target for machine learning to start our investigations in this new domain. One direction could be to include name binding features (following $\lambda$-Prolog~\cite{Miller:2012} or $\alpha$-Prolog~\cite{Cheney:2008}) that make it easier to implement substitution.

Another direction is to test the method on more complex semantic rules. Modular structural operational semantics (MSOS)~\cite{Mosses:2004} gives us hope that it is possible: it expresses the semantics of complex languages in a modular way, which means that the rules do not need to be changed when other rules change. MSOS can be implemented in Prolog.

For a working system we also need some semi-automatic translation from the concrete syntax of the language to abstract syntax. This is a different research problem, but could also be a suitable candidate for ILP. 

Krishnamurthi et al.~\cite{Krishnamurthi:2019} framed the same general problem differently: they assume that we know the core semantics in the form of an abstract language, and we need to learn syntactic transformations in the form of tree transducers that reduce the full language to this core language. They attempted several learning techniques, each with shortcomings, but did not consider ILP, so applying ILP to their problem could be an interesting direction to take.

\paragraph{Acknowledgments}
The authors wish to thank Andrew Cropper, Vaishak Belle, and anonymous reviewers for
comments.  This work was supported by ERC Consolidator
Grant Skye (grant number 682315).

\bibliographystyle{splncs04}
\bibliography{milpls}

\end{document}